\documentclass[amsmath,amssymb,superscriptaddress,a4paper,preprint]{article}

\usepackage{amsmath,amsfonts,amssymb}
\usepackage{graphicx,float}
\usepackage[colorlinks,citecolor=blue,urlcolor=blue]{hyperref}
\usepackage[left=2cm,top=2.5cm, right=2cm, bottom=2cm]{geometry}
\usepackage{subfigure}
\usepackage{caption}
\usepackage{bm}
\usepackage{mathrsfs}
\usepackage{txfonts}
\usepackage[amssymb]{SIunits}
\usepackage{epsfig}
\usepackage{lipsum}
\usepackage{color}
\usepackage{textcomp}
\usepackage{multirow}
\usepackage{multicol}
\usepackage[T1]{fontenc}
\usepackage{times}
\usepackage[utf8]{inputenc}
\usepackage{placeins}
\usepackage[sort,numbers,compress]{natbib}
\usepackage{authblk}
\usepackage{lineno}

\allowdisplaybreaks[2]
\title{Kerr-type nonlinear optical isolators bypassing dynamic reciprocity}

\author[1,2]{Yiqi Hu}
\author[1]{Yihong Qi}
\author[1,2]{Yu You}
\author[1]{Shicheng Zhang}
\author[1]{Gongwei Lin}
\author[1]{Xiaolin Li}
\author[3]{Jiangbin Gong}
\author[1,4]{Shangqing Gong}
\author[1,4]{Yueping Niu\thanks{niuyp@ecust.edu.cn}}

\affil[1]{Department of Physics, East China University of Science and Technology, Shanghai 200237, China}
\affil[2]{School of materials Science and Engineering, East China University of Science and Technology, Shanghai 200237, China}
\affil[3]{Department of Physics, National University of Singapore, 117542, Singapore}
\affil[4]{Shanghai Engineering Research Center of Hierarchical Nanomaterials, Shanghai 200237, China}

\date{}

\begin{document}
\captionsetup[figure]{labelfont={bf},name={Fig.}}
\maketitle

\noindent  Magnetic-free optical isolators are critical components for the realization of integrated optical systems.  The underlying physics of passive nonlinear optical isolators is not solely about breaking the Lorentz reciprocity without requiring any external bias. Indeed, one major obstacle to the operation of Kerr-type nonlinear optical isolators was found to be the so-called dynamic reciprocity, of which the serious outcome is that a backward signal cannot be isolated in the presence of a forward strong signal. In this work, we advocate the novel concept of velocity-selective nonlinearity to bypass such dynamic reciprocity.  Using a proof-of-principle platform with warm rubidium atoms in an asymmetric cavity, we experimentally achieve the isolation of a backward signal in the presence of a strong forward signal, with experimental observations in agreement with our theoretical simulations. This work has thus made one essential step towards functioning Kerr-type passive optical isolators.
\newpage
\onecolumn

Optical isolators are expected to have a wide variety of applications in 
all-optical signal processing. They are non-reciprocal devices that allow light 
to transmit in one direction but completely suppress light propagation in the 
reverse direction. One prerequisite for realizing optical isolators is to break 
the Lorentz reciprocity  
\cite{NP-Jalas-7}.  Magnetic bias may easily break optical reciprocity, but
this is not a desirable solution in integrated photonics. Indeed, a number of 
alternative magnetic-free non-reciprocal schemes based on external electric or laser bias, involving different branches of physics and a various of platforms, have been proposed in the recent decade, including spatio-temporal modulation \cite{NP-Yu-3,PRL-Lira-109,NP-Estep-10}, 
chiral quantum optics \cite{Nat-Lodahl-541,PRA-Xia-90,Sci-Scheucher-354}, 
and optomechanical systems \cite{NP-Shen-10,NC-Ruesink-7}. 

Of particular interest is the use of Kerr-type nonlinear effects in the design 
of optical isolators 
\cite{Sci-Fan-335,OP-Del-5,NC-Mahmoud-6,PRL-Yang-123,PRL-Hanmann-90}.
Kerr-type optical isolators represent a promising route towards passive devices that do not require any form of external bias, with two rather fundamental issues however.  The first is the nontrivial trade-off between the transmission of forward signal and the achievable range of isolation intensity \cite{PRB-Sounas-10}. Fortunately, this complication has been largely overcome by combining multiple nonlinear resonators \cite{NE-Sounas-1,NP-K}. The second and more serious restriction, namely, the dynamic reciprocity, 
was also discovered only several years back \cite{NP-Shi-9}.  Due to dynamic reciprocity,  
Kerr-type nonlinear isolators do not actually operate in the presence of the forward signal. That is, there is no transmission suppression of the backward signal if the forward and backward signals are both injected.  Such an unpleasant feature is inherent to Kerr-type nonlinearity and severely hinders further progresses in theoretical and experimental studies of passive optical isolators.

Our proposal in this work exploits the seemingly unrelated thermal motion of atoms in an atomic medium. 
Random thermal motion of atoms tends to wash out subtle quantum effects and is therefore seen as a destructive nuisance in the quantum control of atoms by light or vice versa. Yet, this is a naive perspective because the skillful use of atomic translational motion in connection with the Doppler effect has made a ground-breaking impact on laser cooling. 
Indeed, by velocity-selective coherent population trapping, Cohen-Tannoudj $ 
et~al. $ first pioneered in achieving laser cooling below the one-photon recoil 
energy \cite{PRL-Aspect-61}.  Most recently, velocity-selective 
electromagnetically induced transparency \cite{NP-Zhang-12,PRL-Xia-121} has been designed to achieve non-reciprocal quantum optics. Here, we advocate the novel concept of  velocity-selective nonlinearity, based on which we demonstrate how passive nonlinear isolators can be realized, thus bypassing the curse of dynamic reciprocity.

\subsection*{Results}

\noindent {\bf Theoretical model.} Most passive nonlinear isolators adopt spatially asymmetric structure together 
with a nonlinear medium.  The dielectric property of the medium depends on the 
intensity of the incident signals. The asymmetric structure is the origin of 
non-reciprocal light propagation because such structure allows to inject the 
forward (backward) signal with high (weak) intensity in the same nonlinear 
medium. Then Kerr-type passive non-reciprocity yields high transmission for 
the forward signal and isolation for the backward signal, as observed in many 
experiments 
\cite{Sci-Fan-335,OP-Del-5,NC-Mahmoud-6,PRL-Yang-123,PRL-Hanmann-90}.
Unfortunately, these progresses only represent an incomplete solution to 
optical isolators due to dynamic reciprocity discovered in 
Ref.~\cite{NP-Shi-9}.  That is, the isolation of the backward signal ceases to 
work if the forward signal is also present in the system.  To enlighten on 
this,  Ref.~\cite{NP-Shi-9} considered the transmission of a weak additional 
wave $ \mathbf{E}_{s} $ (with frequency $ \omega_s $) in either the forward or 
backward direction,  in the presence of a high-intensity forward wave $ 
\mathbf{E}_{0} $  (with frequency $ \omega_0 $).  The propagation equation of 
the weak additional wave becomes
\begin{equation}
\begin{aligned}
&\bigtriangledown \times \bigtriangledown \times \mathbf{E}_{s} - 
\omega_{s}^{2} 
\mu_{0}[ \mathbf{\epsilon} (\omega_{s},\mathbf{r})\\
&+6 \mathbf{\epsilon}_0 \chi^{(3)}(\omega_{0},\omega_{s},\mathbf{r}) 
|\mathbf{E}_{0}|^2]\mathbf{E}_{s}=0,
\end{aligned}
\end{equation}
with the dielectric function modified as
\begin{equation}
\textbf{$ \epsilon 
	$}_{s}\to \textbf{$ \epsilon 
	$} (\omega_{s},\textbf{r})+6 \textbf{$ \epsilon $}_0 
\chi^{(3)}(\omega_{0},\omega_{s},\textbf{r}) |\textbf{E}_{0}|^2,
\end{equation}
where $\chi^{(3)}$ is the standard Kerr nonlinearity coefficient. 
For general Kerr-type nonlinear systems, the modified dielectric function $ 
\epsilon_{s} $ is apparently a time-independent scalar and behaves precisely 
the same for forward and backward signals.
Equation (1) thus predicts reciprocal light propagation if a high-intensity forward 
signal is already present \cite{NP-Shi-9}, indicating that the weak additional 
wave in the backward direction must have high 
transmission as what can be expected if we add a weak signal in the forward direction.   

\begin{figure*}[]
	\centering
	\includegraphics[width=17cm]{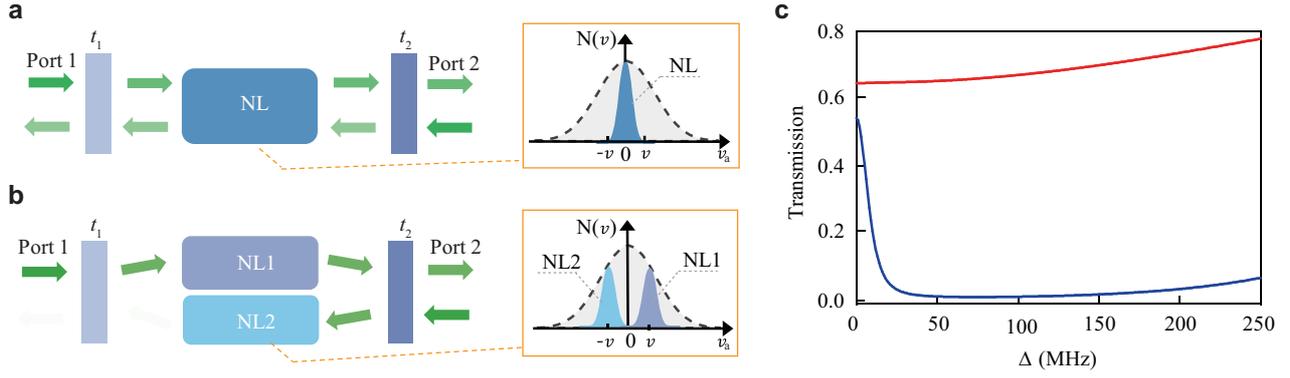}
	\caption{\textbf{Operating principle for bypassing dynamic 
			reciprocity.} \textbf{a}, A two-port nonlinear isolator with 
		spatially asymmetric structure. Without {\it velocity-selective nonlinearity}, the forward and backward signals will interact with 
		the same group of atoms near zero velocity (labeled NL). The large 
		intensity signal from the forward direction (left to right) will ``open 
		a channel" for transmission of the weak signal from backward (right 
		to left). Then the backward signal cannot be isolated if the 
		forward signal is present at the same time. This so-called ``dynamic 
		reciprocity” discovered in Ref.~\cite{NP-Shi-9} causes serious 
		limitation for passive nonlinear isolators. 
		\textbf{b}, With {\it velocity-selective nonlinearity}, the forward and 
		backward signals selectively interact with two sub-ensembles of 
		atoms with different peak velocities $v$ and $-v$ (labeled NL1 and NL2). The forward 
		large-intensity signal can only open the transmission channel in 
		NL1 and will not affect the backward weak signal propagating in NL2. The dynamic reciprocity is therefore bypassed. 
		\textbf{c},  Theoretical simulated transmission of the forward (red line) and backward (blue line) signals when they are injected simultaneously. }
	\label{fig1}
\end{figure*}

Bypassing or overcoming dynamic reciprocity is hence a key scientific and engineering problem in designing passive optical isolators. One solution is to go beyond Kerr-type nonlinearity via four-wave mixing \cite{NC-Hua-7,IEEE-Wang-28}, which nevertheless is active and needs external laser bias. Here we offer an innovative route to bypass (but not break) dynamic reciprocity, in the context of a nonlinear and warm atomic medium. 

Consider now the thermal motion of atoms in a nonlinear medium. The atom velocity moving along the forward laser beam is assumed to be $v$ below. Accounting only for the longitudinal Doppler effect along the laser beam direction (the transverse Doppler effect can be ignored), the laser frequency $\omega$ is thus shifted to $ \omega-kv $ or  $ \omega+kv $ for the forward or backward propagation cases, with $k$ being the laser wavevector. The above dielectric function for the weak additional wave propagating forward (+) and backward (-) should then be modified and becomes
\begin{equation}
\textbf{$ \epsilon 
	$}_{s\pm}\to\textbf{$ \epsilon 
	$} (\omega_{s}\mp kv,\textbf{r})+6 
\textbf{$ \epsilon $}_0 
\chi^{(3)}(\omega_{0}-kv,\omega_{s}\mp kv,\textbf{r}) 
|\textbf{E}_{0}|^2.
\end{equation} 
As such, so long as $v$ is appreciably nonzero, the modified dielectric function  $\textbf{$\epsilon$}_{s+}=\textbf{$\epsilon$} 
(\omega_{s}-kv,\textbf{r})+6 \textbf{$\epsilon$}_0  \chi^{(3)}(\omega_{0}-kv,\omega_{s}-kv,\textbf{r}) |\textbf{E}_{0}|^2$ for the additional forward signal
cannot be the same as that  $\textbf{$\epsilon$}_{s-}=\textbf{$\epsilon$} 
(\omega_{s}+kv,\textbf{r})+6 \textbf{$\epsilon$}_0  \chi^{(3)}(\omega_{0}-kv,\omega_{s}+kv,\textbf{r}) |\textbf{E}_{0}|^2$ for the additional backward signal. The forward and backward weak signals can therefore have completely different nonlinear optical responses. Indeed, as schematically depicted in Fig.~1a, in the absence of atomic thermal motion, the forward and backward weak signals interact with the same group of atoms (nonlinear medium), of which the atom velocities are peaked at zero. Consequently the forward high-intensity signal ``opens a channel" through Kerr-type nonlinearity, yielding the same transmission for any additional signal in both forward and backward directions, just as discussed in Ref.~\cite{NP-Shi-9}. However, upon considering random thermal motion of atoms in the medium, a new physical mechanism is emerging. To see this, note first that the Maxwell-Boltzmann velocity distribution of the moving atoms are symmetric with respect to $v\rightarrow -v$.  So if the forward signal interacts resonantly with moving atoms whose velocities are peaked at $v$, the backward signal must interact resonantly with a different group of atoms with velocities peaked at $-v$ (see Fig. 1b). Because the forward high-intensity field only shares the same Doppler effects with the forward weak signal but not with the backward one, it opens up a channel through the Kerr-type nonlinearity for the forward weak signal, but does not affect the transmission of the backward one.  We refer to such independent nonlinear optical responses for the forward and backward signals as {\it velocity-selective nonlinearity}, as a consequence of Doppler effects effectively dividing one single nonlinear medium into two. This way,  we have bypassed dynamic reciprocity and then the isolation of a backward weak signal is achievable even in the presence of a forward high-intensity signal.

\begin{figure*}[]
	\centering
	\includegraphics[width=17cm]{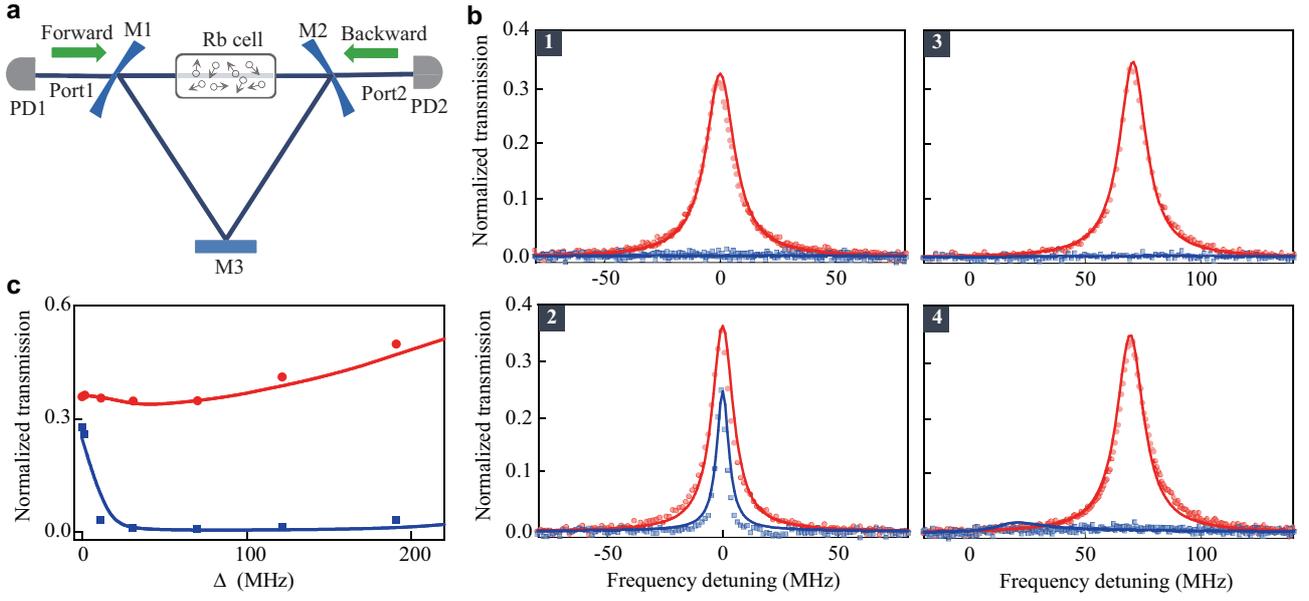}
	\caption{ \textbf{Experimental setup and results.} \textbf{a}, Schematic of 
		experimental setup: M1, M2 and M3 are cavity mirrors and form an 
		asymmetric cavity. Rb atoms are used as the nonlinear medium. The 
		forward and backward signals are detected by photodetector 2 (PD2) and 
		1 (PD1), respectively.  
		\textbf{b}, Normalized transmission of the forward  (red dot) and backward 
		(blue square) signals. Panel 1, 2: The forward and backward signals are 
		incident separately (Panel 1) or simultaneously (Panel 2) without the 
		mechanism of velocity-selective nonlinearity. Panel 3, 4: The forward 
		and backward signals are incident separately (Panel 3) or 
		simultaneously (Panel 4) with our proposed velocity-selective 
		nonlinearity in action. The forward and backward signal intensity are 
		fixed at 0.6~mW. 
		\textbf{c}, Normalized transmission of forward and backward 
		signals as a function of $ \Delta $ when they are injected simultaneously. Results from theoretical simulations 
		are presented here as solid lines.}
	\label{fig2}
\end{figure*}

The above principle motivated us to consider saturated absorption of two-level atoms (with resonant frequency $ \omega_a $) at room temperature as Kerr-type nonlinearity to demonstrate the bypassing of dynamic reciprocity. For atoms with velocity $v $, the resonant interaction occurs at $ \omega_s =\omega_a+kv $  for the forward signal because of the Doppler shift. This being the case, the selection of sub-ensembles of atoms at different characteristic velocities can be achieved straightforwardly by the interplay of resonant interaction and Doppler effect. Indeed, by adjusting the frequency detuning of the signal field, i.e., $\Delta =\omega_s - \omega_a$,  one can now tune the peak velocity of the first sub-ensemble atoms under which saturated absorption for the forward signal occurs, with $v= \Delta/k$.  Accordingly, the peak velocity of the second sub-ensemble of atoms on resonance with the backward signal is tuned to be  $-v=\Delta/(-k)$.  As shown in Fig.~1c, signal transmission is theoretically investigated as a function of $ \Delta $ under the simultaneous action of strong forward and weak backward signals (with details provided in Supplementary Information). For cases with small $\Delta$, both forward and backward signal have high transmission due to the overlap of the two atomic sub-ensembles, thereby possessing dynamic reciprocity. With the increase of $ \Delta $, the forward signal maintains a high transmission whereas the backward signal is suppressed. As elaborated above, this is because the two atomic sub-ensembles selected by forward and backward signals are separated further apart with increasing $\Delta$, and hence velocity-selective nonlinearity comes into play. Theoretical results in Fig.~1c indicate that over a wide frequency range, dynamic reciprocity can be bypassed.
\vspace{6pt}

\begin{figure*}[]
	\centering
	\includegraphics[width=12.1 cm]{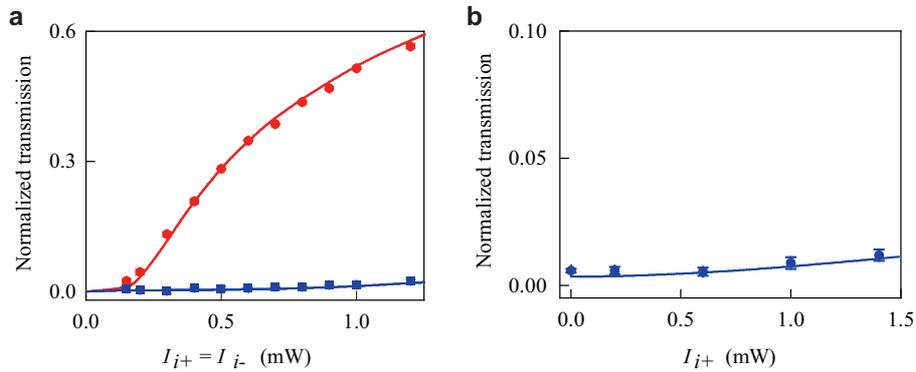}
	\caption{\textbf{Experimental observation of the dependence of signal transmission on light intensity, corresponding to the case of Panel 4 in Fig. 2b.} \textbf{a}, Normalized transmission of the forward (red dot) and backward (blue square) signal. \textbf{b}, Normalized transmission of the backward signal with its intensity fixed at 0.6~mW. The error bars show the standard deviation. Results from theoretical simulations are presented as solid lines.}
	\label{fig3}
\end{figure*}

\noindent {\bf Experimental realization.} To verify our concept, we now report a proof-of-principle experiment. Warm $ ^{87} $Rb  atoms of two energy levels are used as the nonlinear medium. As shown in Fig.~2a, a ring cavity with asymmetric cavity mirrors M1 and M2 is constructed to obtain different intensity of intracavity signals with the same intensity inputs from opposite directions. As a result, the forward signal propagates with high intracavity intensity interacting nonlinearly with the atomic medium and experiences saturated absorption; whereas the backward signal is very weak and so it is strongly absorbed (experimental details can be found in methods).

To benchmark our system in view of dynamic reciprocity, we first work with a peculiar parameter regime that can effectively turn off the mechanism of velocity-selective nonlinearity. To that end, we consider a signal on resonance with the atomic transition ($ \Delta = 0 $). Our experimental results are shown in Panel~1 and 2 of Fig.~2b. Panel~1 displays a high transmission for the forward signal and excellent isolation for the backward signal when they are incident separately. This non-reciprocal transmission with only one signal field demonstrates the basic design of passive isolators. Next we inject the 
forward and backward signals into the system simultaneously. This time, the 
backward signal, which can be well isolated in the absence of the 
forward signal, is seen to have a significant transmission analogous to that of 
the forward signal (Panel~2). Evidently then, the high-intensity 
forward signal incident here ``opens a channel” for the backward signal. The 
dynamic reciprocity is thus recovered as expected. Indeed, without 
velocity-selective nonlinearity, both the forward and backward signals interact 
with essentially the same nonlinear medium and dynamic reciprocity must set an 
obstacle to optical isolation. 

We now proceed with a new set of experiments, by introducing a considerable 
detuning  $ \Delta $. According to our simulation results shown in Fig.~1c, $ \Delta=70 $~MHz is chosen.
Experimental results are shown in Panel~3 and 4 of Fig.~2b. When the forward 
and backward signals are incident separately (Panel~3), 
the forward one has high transmission whereas the backward is isolated, which 
is essentially the same as Panel~1. Most significantly, when both 
the forward and backward signals are incident simultaneously, the backward 
signal now stays being isolated (Panel~4). This is in sharp contrast 
to Panel~2 and clearly signifies that the dynamic reciprocity is now 
bypassed. To digest this bypass, consider Doppler effects once again on the modified dilectric function. The forward and backward signals with detuning $ \Delta $ = 70~MHz interact with atoms of velocities peaked at $ v = \Delta/k $ and $ -v = \Delta/(-k) $ respectively. As such, whether the forward and backward signals are incident separately or simultaneously, they interact with two different sub-ensembles of the same nonlinear medium. This underlying division of one single nonlinear medium into effectively two is an engineering gift from nature.
Consequently, the high-intensity forward signal, though reaching saturated absorption for one sub-ensemble, is not capable  to ``open the channel” for the backward signal interacting with the other sub-ensemble of atoms. Our results here constitute the first experimental demonstration of a Kerr-type passive optical isolator that can bypass dynamic reciprocity. In addition, we also 
measured the non-reciprocal transmission at different signal frequency detuning 
when the forward and backward signals are incident simultaneously. As shown in Fig.~2c, over a wide frequency range, the backward signal can be isolated well. It should be also noted that our experimental results in Fig.~2 are in excellent agreement with our theoretical simulations. 

To further characterize the performance of our prototype nonlinear optical isolator, we have also measured its intensity response with the mechanism of velocity-selective nonlinearity switched on. Equal intensity ranging from 0.15~mW to 1.2~mW for the simultaneously incident forward and backward signals is used. As Fig. 3a shows,  with the increase of signal intensity, saturated absorption occurs, leading to the increase of forward transmission. The backward signal is isolated well for the entire intensity range. In addition, for a fixed backward signal intensity $ I_{i-} $ = 0.6~mW, it is isolated well as the forward signal intensity varies from 0 to 1.4~mW (Fig.~3b). This demonstrates that the isolation of backward signal is not affected even if the injected forward signal becomes much more intense than the backward signal. Our experimental observations on the performance of isolation also fairly agree with our theoretical results. 

\subsection*{Conclusion}

\noindent In summary, velocity-selective nonlinearity has been exploited to successfully  bypass dynamic reciprocity as a fundamental obstacle towards the realization of Kerr-type passive optical isolators.  The forward and backward signals selectively interact with two different sub-ensembles of  warm atoms with different characteristic velocities. This selection or division of one nonlinear medium 
into effectively two is spontaneously done by Doppler effects in laser-atom 
interaction.  Proof-of-principle experiments using saturated absorption have 
clearly demonstrated the isolation of a backward signal in the presence of a 
strong forward signal.  Our approach can be further integrated into optical 
chip-based cavities 
\cite{Nat-Vahala-424,NJP-Ritter-18,LPR-Stern-10,OL-Jones-41}, thus opening a 
door for realizing integrated passive optical non-reciprocal devices.

\section*{Methods}
\small
\noindent {\bf Cavity and atom parameters.}
The ring cavity is composed of three mirrors: two concave mirrors M1 and M2 (curvature radius $40~\centi\meter$) with different reflectivity of $99\%$ and $99.9\%$; a flat mirror M3 with reflectivity of $99.95\%$. The cavity mirror M3 is mounted on a piezoelectric transducer (PZT) for cavity frequency tuning. The optical length of the cavity is about 
$40~\centi\meter$, corresponding to a free spectral range (FSR) about 
$750~\mega\hertz$. The finesse of the cavity with $5~\centi\meter$ long Rb vapor cell in is about $100$. The Rb vapor cell is wrapped in $\mu$ metal and keeps at  $50~\celsius$. The two levels we used are $5^2S_{1/2}, F=1$ and $5^2P_{1/2}, F'=2$ of $^{87}$Rb D2 line.

\noindent {\bf Transmission measurement.}
In our scheme, the mechanism of velocity-selective nonlinearity is turned off or on by selecting signal field resonance ($\Delta = 0$) or detuning ($\Delta \neq 0$). This can be achieved by adjusting the cavity frequency using PZT in the experiment. Signals that resonate with the cavity can enter the cavity (intracavity signals) and interact with the atomic nonlinear medium. We scan the frequency of the signal and then the intracavity signal can be tuned to be resonant or have certain detuning with the atomic transition by controlling the PZT. During this process, the frequency of the intracavity signal is calibrated by saturation absorption spectroscopy (SAS) setup. The experimental data of the transmission shown in Fig.~2 and~3 are normalized by the cavity transmission away from the atomic resonance.

\section*{Acknowledgements}
This work was supported by the National Natural Science Foundation of China  
(grant nos. 11974109, 11774089, 11874146, 11674094, 61835013) and the Shanghai Natural Science Foundation (grant nos. 20YF1410800, 17ZR1442700, 18ZR1410500). 

\section*{Author contributions}
Y.N., Y.H. and Y.Q. contributed to the original idea, and Y.N. conceived the manuscript. Y.H. carried out the theoretical simulation and conducted the experiment. Y.H. and Y.Q. contributed equally to this work. J.G. partly contributed to refining the idea and also finalized this manuscript.  S.G. supervised the whole project. All authors participated in the discussion and writing of the manuscript.

\section*{Additional information}

\noindent \textbf{Competing interests:} The authors declare no competing interest.

\end{document}